\documentclass[mathpazo]{cicp}

\usepackage{mathrsfs,amsmath,amssymb}
\usepackage[numbers]{natbib}
\usepackage[T1]{fontenc}
\usepackage{graphicx}
\usepackage{subcaption}
\usepackage{xcolor}
\usepackage{tabularx}
\usepackage{booktabs}
\usepackage{enumerate}
\begin{document}

\title{ A simulation approach including under-resolved scales for multi-component fluid flows in multi-scale porous structures}


\author[O.~Author]{Hiroshi Otomo\corrauth , Rafael Salazar-Tio, Jingjing Yang, Hongli Fan, Andrew Fager, Bernd Crouse, Raoyang Zhang, Hudong Chen}

\address{Dassault Syst\'{e}mes, 185 Wyman Street, Waltham, MA 02451, USA}
\email{\tt Hiroshi.Otomo@3ds.com} %(O.~Author)
%address{School of Mathematical Sciences, Beijing Normal University,
%Beijing 100875, P.R. China}
%\email{{\tt author@email} (O.~Author)}

% \date{\today}

\begin{abstract}
In this study, we develop computational models and methodology for accurate multi-component-flow simulation in under-resolved multi-scale porous structures \cite{2020_Hotomo_patent}. 
It is generally impractical to fully resolve the flow in porous structures with large length-scale difference due to tremendously high computational expense. The flow contributions from under-resolved scales need to be accounted with proper physics modeling as well as simulation processes. Using pre-computed physical properties such as the absolute permeability, $K_0$, the capillary-pressures-saturation curve, and the relative permeability, $K_r$, in typical resolved porous structures, local fluid force is conjectured and applied to simulation in the under-resolved regions which are represented by porous media. By doing so, accurate simulation of flow in multi-scale porous structures becomes feasible.

In order to check accuracy and robustness of this method, a set of benchmark test cases are performed for both single-component and multi-component flows in artificially constructed multi-scale porous structures, and simulation results are compared with analytic solutions and/or results with much finer resolution resolving the porous structures. Quantitatively consistent results are obtained with proper input of $K_0$, capillary pressure, and $K_r$ in all tested cases.
Specifically, imbibition patterns, entry pressure, residual component’s patterns, and the absolute/relative permeability are accurately captured with this approach. 
\end{abstract}

% \pacs{Valid PACS appear here}% PACS, the Physics and Astronomy
% \keywords{Suggested keywords}%Use showkeys class option if keyword
                              
                              %%%%% AMS/PACs/Keywords %%%%%%%%%%%
%\pac{}

%%%% maketitle %%%%%
                              
\maketitle
\small

{\bfseries This is an accepted manuscript version of the paper published in Communications in Computational Physics, Vol. 33(1)189, , 2023. The final published version is available at https://doi.org/10.4208/cicp.OA-2022-0037.}

%%%%%%%%%%%%%%%%%%%%%%%%%%%%%%%%%%%%%%%%%%%%%%%%%%%%%%%%%%%%%%%%%%%%%%%%%%%%%%%%%%%%%%%%%%%%%%%%%%%%%%%%%%%%%%%%%%%%%%%%%%%%%
%
%Introduction
%
%%%%%%%%%%%%%%%%%%%%%%%%%%%%%%%%%%%%%%%%%%%%%%%%%%%%%%%%%%%%%%%%%%%%%%%%%%%%%%%%%%%%%%%%%%%%%%%%%%%%%%%%%%%%%%%%%%%%%%%%%%%%%

%%%%%%%%%%%%%%%%%%%%
\section{Introduction}
%%%%%%%%%%%%%%%%%%%%

Numerical simulation of multi-component fluid flows in porous regions with complex solid structures are of great importance in many industrial applications, for example 
 enhanced oil recovery including carbon-dioxide injection, capture and storage \cite{2021_Han,2021_Shirbazo},
 water/air flow in gas diffusion layers of the fuel cells, \cite{2017_Xu,2019_Xiao,2020_Mortazavi},  
 in-situ copper mining by leaching \cite{2005_Cariaga},
 and sophisticated personal protective equipment \cite{2021_Tcharkhtchi}. 
In order to achieve high fidelity simulation, it is crucial to fully resolve complex solid boundaries. In most of simulation cases, however, fully resolving all details of a multi-scale porous structure is prohibited due to limited machine power as well as immaturity of computational models and algorithms, although such complex structures are frequently observed in the nature.

\begin{figure*}[htbp]
  \begin{center}
    \includegraphics[clip, width=10 cm]{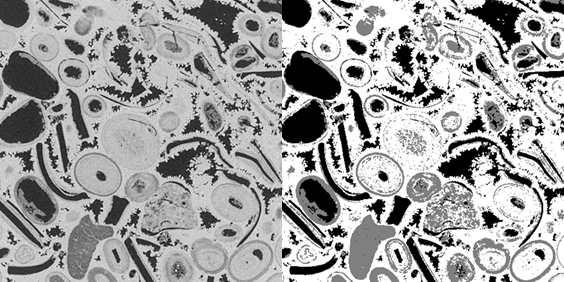}
    \caption{A cross section pictures of a Carbonate rock. An original scanned picture (left) and a segmented picture with small-scale porous structure marked in grey (right) are shown. }
    \label{fig:exampled_rock_pictures}
  \end{center}
\end{figure*}

Here is one example from the oil\&gas industry application.  Fig.~\ref{fig:exampled_rock_pictures} shows a typical cross section of a Carbonate-rock sample that has  porous structures with multiple different scales \cite{2016_Bultreys}. The left picture is one slice of the original micro-tomography scanned image, and the right is its segmented image with small-scale porous structures marked in grey. The length-scale difference between black and grey structures is approximately 10 times. 
The small-scale porous regions in grey could significantly impact the flow behavior in the large scale because it can contribute to  the connectivity among larger-scale pores and can lead to high capillary forces and  variable flow effective resistivity, for instance. Therefore it is necessary to properly take their contributions into account. 
However, resolving all of such small-scale details requires extremely fine resolutions, that results in a tremendously expensive simulation. The cost could increase by 10s of thousand times, compared to the unresolved case ignoring the  small porous structures contributions from the grey regions, because of the increased number of three-dimensional cells and the reduction of time increments. Therefore such fully resolved simulations are impractical in industrial applications. 

%+++++ Figure (Result: Capillary fingering) ++++++
%+++++ Figure end +++++

In many previous studies \cite{1998_Freed,2002_Martys,2015_Ginzburg,2017_Li,2018_Zhu,2019_Fager,2020_Zhang}, the viscous force from under-resolved porous media (PM) was modeled  by a resistance term, like in the Brinkman equation, using a pre-computed permeability in the resolved-PM at finer resolutions. They are, however, mainly focusing on the single-component fluid flow and have not extended to the multi-component fluid flow.
Few recent studies \cite{2020_Dinarievi,2020_Suhrer} discuss about the multi-component fluid flow in the multi-scale porous structures.
In reference \cite{2020_Dinarievi}, relative permeability from under-resolved  PM regions was computed by solving the transport equation for the total energy, Helmholtz free and kinetic energy, without referring to physical properties such as the capillary-pressure curves. 
In reference \cite{2020_Suhrer}, the pre-computed capillary pressure and the  effective flow resistance in the resolved-PM are applied for the force balance analysis at the capillary equilibrium state using recursive methods.
Due to the sensitivity of solutions for the multi-component fluid flow on the initial flow conditions,  it is desirable to solve the unsteady fluid-dynamics equation  even for capturing the steady state.
In this study, we solve equations based on the unsteady fluid dynamics following conditions  similar in laboratory experiments. 
Besides the accuracy, this approach allows us to check the transition of the components' distribution and flow pattern.
Moreover, handling both of the input and output data for the PM model by a single solver can contribute to the  simulation robustness in a significant way. 

 Here, we propose a solution based on the lattice Boltzmann method, although the methodology itself is not limited on it. 
 In the proposed workflow, computational models are implemented to account for effects of fluid flow in the under-resolved regions.
This approach should be applicable for various engineering cases of multi-scale porous systems.

This manuscript is organized as follows. 
A basic formalism of the lattice Boltzmann method (LBM) for the multi-component fluid flow is introduced in Sec.~\ref{sec:LB_formalism}. 
Proposed workflow and numerical models for treating multi-scale structures are presented in Sec.~\ref{sec:LB_formalism_PM}. 
In Sec.~\ref{sec:validation},  detailed settings and results in a set of benchmark test cases for single-component and multi-component fluid flows are shown. Finally,
in Sec.~\ref{sec:summary}, findings in this study are summarized.

%%%%%%%%%%%%%%%%%%%%
\section{Lattice Boltzmann models for immiscible fluids}
\label{sec:LB_formalism}
%%%%%%%%%%%%%%%%%%%%

Lattice Boltzmann models for immiscible fluids are introduced in this section, which are based on the Shan-Chen model \cite{1993_Xiaowen,1994_Xiaowen} and its recent advancements \cite{2006_Chen, 2006_Zhang, 2006_Latt, 2006_Xiaowen_JFM, 2012_Qli}. 
The lattice Boltzmann (LB) equation for multi-component fluid is: 
\begin{equation}
\label{eq:basic_LBM_eq}
 f_{i}^{\alpha} \left( \vec{x}+\vec{c}_{i} \Delta t, t+\Delta t \right) - f_{i}^{\alpha} \left( \vec{x} , t \right)
= \mathcal{C}_{i}^{\: \alpha} + \mathcal{F}_{i}^{\alpha},
\end{equation}
	where $f_i^{\alpha}$ is the density distribution function of fluid component $\alpha$ and $\vec{c}_{i}$ is the discrete particle velocity.  In this study, binary mixture of immiscible fluids, such as water and oil, is considered for simplicity, namely $\alpha=\left\{water (w), oil (o) \right\}$, although the framework can be easily extended to arbitrary number of components. 
The D3Q19 \cite{1992_Qian} lattice model with the fourth order lattice isotropy is employed.

The simplest form of the collision operator $\mathcal{C}_{i}^{\: \alpha}$ is the Bhatnagar-Gross-Krook type,

\begin{equation}
\label{col_operator}
\mathcal{C}_{i}^{\: \alpha} = -\frac{1}{\tau_{mix}}(f_{i}^{\alpha} - f_{i}^{eq, \alpha}),
\end{equation}
where $f_{i}^{eq, \alpha}$ is the equilibrium distribution function for the Stokes flow with the third order expansion in $\vec{u}$ ,
\begin{equation}
\label{eq:feq}
f_{i}^{eq, \alpha} = \rho_{\alpha} w_{i} \left[ 1 + \frac{\vec{c}_{i} \cdot \vec{u}}{T_0}  +  \frac{\left( \vec{c}_{i} \cdot \vec{u} \right)^3}{6T_0^3} -  \frac{\vec{c}_{i} \cdot \vec{u}}{2T_0^2}\vec{u}^2 \right],
\end{equation}
where $T_0 = 1/3$ and $w_i$ denote the lattice temperature and isotropic weights in D3Q19, respectively. 
The density of the component $\alpha$, $\rho_{\alpha}$, and the mixture flow velocity, $\vec{u}$, are defined as,
\begin{eqnarray} 
\rho_{\alpha} = \sum_i f_i^{\alpha}, \; \; \;
\rho = \sum_{\alpha} \rho_{\alpha} = \sum_{\alpha} \sum_i f_i^{\alpha}, \; \; \;
\vec{u^{\alpha}} = \sum_i \vec{c}_{i} \cdot f_i^{\alpha} / \rho_{\alpha}, \; \; \;
\vec{u} = \sum_{\alpha} \sum_i \vec{c}_{i} \cdot f_i^{\alpha} / \rho.
\end{eqnarray}
The relaxation time $\tau_{mix}$ in Eq.~(\ref{col_operator}) relates to the kinematic viscosity of the mixture of components, $\nu_{mix}$, as
\begin{equation}
\tau_{mix} = \left( \nu_{mix} / T_0  \right) + 1/2,
\end{equation}
\begin{equation}
\label{mix_nu}
\nu_{mix}  = \left( \rho_{w} \nu_{w}+\rho_{o}\nu_{o} \right)/ \left(\rho_{w}+\rho_{o}\right).
\end{equation}
Following the conventional way \cite{1993_Xiaowen,1994_Xiaowen}, the inter-component force, $\vec{F}^{\alpha, \beta}$, between component $\alpha$ and $\beta$ is defined as, 
\begin{equation}
\label{eq:comp_interaction}
\vec{F}^{\alpha, \beta} \left( \vec{x} \right) = G \rho_{\alpha} \left( \vec{x} \right) \sum_{i} w_{i} \vec{c}_{i} \rho_{\beta} \left(  \vec{x}+ \vec{c}_{i} \Delta t \right),
\end{equation} 
for $\alpha \ne \beta$, and  $\vec{F}^{\alpha, \beta} \left( \vec{x} \right) =0$ for $\alpha=\beta$. When the interaction strength $G$ is negative, a repulsive force acts between components and yields a phase separation.
Following reference \cite{2012_Qli}, this inter-component force is implemented in the forcing term $\mathcal{F}_{i}^{\alpha}$ in Eq.~(\ref{eq:basic_LBM_eq}).
The acceleration of the component $\alpha$, $\vec{g}_{\alpha}$, originated from $\vec{F}^{\alpha, \beta}$ is defined by $\vec{g}_{\alpha}= \sum_{\beta} \vec{F}^{\alpha, \beta} / \rho_{\alpha}$. The resulting fluid velocity $\vec{u}_{F}$ is defined as the velocity averaged over pre- and post- collision steps and written as,
\begin{eqnarray}
\vec{u}_{F}= \vec{u}+\vec{g}\Delta t /2,  \; \; \; \; \;
\vec{g} = {\sum_{\alpha} \vec{g}_{\alpha}\rho_{\alpha}} / \rho.
\end{eqnarray}
In what follows, this quantity $\vec{u}_{F}$ is called simply \emph{velocity}.

For enhancing stability and accuracy when $\tau_{mix}$ is not close to 1, a regularized collision operator is used, as described below. 
Rearranging Eq.~(\ref{eq:basic_LBM_eq}), one obtains,
\begin{equation}
\label{eq:LBM_BGK}
 f_{i}^{\alpha} \left( \vec{x}+\vec{c}_{i} \Delta t, t+\Delta t \right) =
 f_{i}^{eq, \alpha} + \left( 1 -
 \frac{1}{ \tau_{mix} }  \right) f_{i}^{' \alpha} + \mathcal{F}_{i}^{\alpha},
\end{equation}
where the function $f_{i}^{'\alpha}$ is the nonequilibrium particle distribution for each fluid component. 
If $f_{i}^{'\alpha}$ takes the standard BGK form $f_{i}^{\alpha}-f_{i}^{eq, \alpha}$ and $\tau_{mix}$ is away from 1, one suffers from the instability caused by unphysical noise and numerical artifacts of the LB model.
To address this issue, a collision procedure regarding $f_{i}^{'\alpha}$ is regulated by,
\begin{equation}
\label{eq:Regualize}
f_{i}^{' \alpha}=\Phi^{\alpha}:\Pi^{\alpha}.
\end{equation}
Here $\Phi$ is a regularization operator that uses Hermite polynomials and $\Pi^{\alpha}$ is the nonequilibrium part of the momentum flux. The basic concept of regularized collision procedure can be found in \cite{2006_Chen, 2006_Zhang, 2006_Latt, 2006_Xiaowen_JFM,1997_Chen,2013_Chen_patent,2020_Chen}. 

% 
% Boundary condition
%

For accurate noslip wall boundary condition on arbitrary geometries, an extension of the volumetric boundary condition proposed by Chen et al in 1998 \cite{1998_Chen,2009_Leo,2004_Yanbing,2006_Fan} is employed. 
In this method, after boundary surfaces are discretized into linear surface facets in two dimension or triangular polygons in three dimension, the incoming and outgoing particles based on those facets or polygons are computed in a volumetric way obeying the conservation laws.  This method is generalized for BC on arbitrary geometry, and it has been studied extensively. More details can be found in \cite{1998_Chen}. 
In order to mitigate numerical smearing in near surface region, especially when physical viscosity is small and the resolution is coarse, a surface scattering model presented in \cite{2009_Leo} is useful.

For the surface wetting conditions, the inter-component interaction force in Eq.~(\ref{eq:comp_interaction}) is extended to the interaction force between wall and fluid particles, $\vec{F}_{w}^{\alpha,  \beta}$, as,
\begin{eqnarray} 
\label{eq:wall_interaction}
\vec{F}_{w}^{\alpha, \beta} \left( \vec{x} \right) = G \rho_{\alpha} \left( \vec{x} \right) \sum_{i} w_{i} \vec{c}_{i} \rho^{'}_{\beta} \left(  \vec{x}+ \vec{c}_{i} \Delta t \right) ,  
\end{eqnarray}
for $\alpha \ne \beta$, and $\vec{F}_{w}^{\alpha, \beta} \left( \vec{x} \right) =0$ for $\alpha = \beta$ where $\rho^{'}_{\beta}$ is constructed  by a fluid part and a solid part $\rho_{\beta}^s$ in a volumetric way so that $\partial \rho_{\beta} / \partial n$ is close to zero \cite{1998_Chen}. This volumetric wettability scheme has a sufficient isotropy on complex geometries \cite{2015_Otomo,2016_Otomo, 2018_Otomo}. 
The wall potential for components, $\rho_{w}^s$ and $\rho_{o}^s$,  is defined as 
\begin{eqnarray} 
\label{wall_potential_def}
\rho_{w}^s=-\rho_0 \rho^s \Theta (-\rho^s), \; \; \: \: \: \rho_{o}^s=\rho_0 \rho^s \Theta (\rho^s),
\end{eqnarray}
using a single parameter $\rho^s$ where $\Theta$ is the Heaviside function and $\rho_0$ is 1.0.

%%%%%%%%%%%%%%%%%%%%
\section{Numerical models and workflow for multi-scale porous structures}
\label{sec:LB_formalism_PM}
%%%%%%%%%%%%%%%%%%%%

In a multi-scale fluid-flow simulation at a certain resolution level, flow contributions from under-resolved porous regions are properly taken into account by applying numerical models at each site using local information of the geometry and fluids. 
The models reproduce proper forces acting on the fluids such as viscous, pressure, and capillary forces, using local representative physical properties such as  absolute permeability $K_0$, relative permeability $K_r^{\alpha}$ and capillary-pressure-saturation curves $P_C - S_w$ where $\alpha$ is an index for the fluid component. These physical properties are pre-computed via fluid-flow simulations in which tiny subdomains of the representative small-scale porous regions are fully resolved. Once they are done, the results are stored in a library.
Each set of physical properties represents a flow type for a particular under-resolved porous structure type. 
In each under-resolved region, a set of physical properties of the porous type is picked from the library and assigned for modelling purposes. By taking into account the local porous geometry information including porosity $\phi$ (the ratio of fluid volume to total volume) and directionality of the structure, the local under-resolved PM flow behavior can be properly reproduced.

\begin{figure*}[htbp]
  \begin{center}
    \includegraphics[clip, width=13 cm]{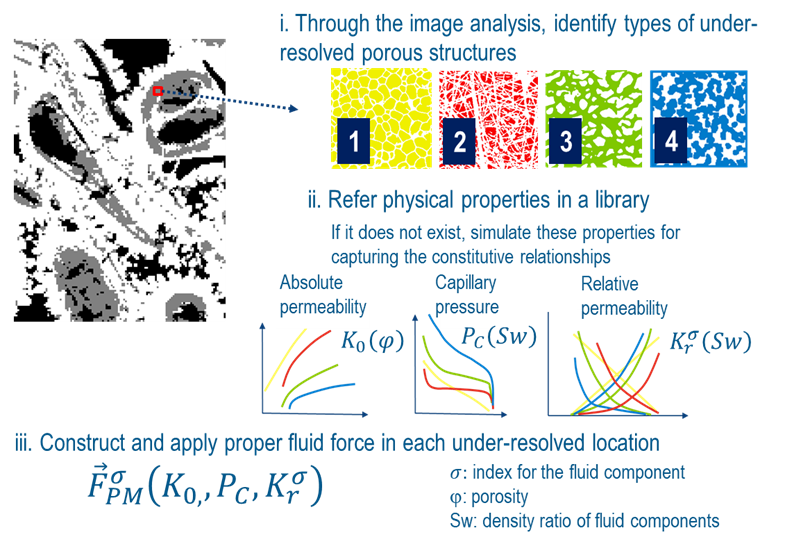}
    \caption{Steps in the procedure }
    \label{fig:steps_procedure}
  \end{center}
\end{figure*}

 The workflow is summarized in Fig.~\ref{fig:steps_procedure} as follows,

\begin{enumerate}[i]
\item	Conduct geometrical analysis of a typical porous structure with scanned images and identify types of under-resolved porous structures. \cite{2021_AFager}
\item	Define representative flow models for each under-resolved region, checking existing sets of physical properties in the library. If the corresponding set of physical properties already exists, pick it up from the library. If not, conduct a fully resolved simulation in a representative subdomain in such under-resolved region, compute the new set of physical properties including absolute permeability, relative permeability, and capillary-pressure-saturation curves for this particular type of porous structure, and add it to the library  utilized for the multi-scale simulation. 
\item	Construct and apply fluid forces at each under-resolved site using the constitutive relationships according to the local geometry information and physical properties. This force corresponds to viscous, pressure and capillary forces from the under-resolved solid structure.
\end{enumerate}

%+++++ Figure (Result: Capillary fingering) ++++++
%+++++ Figure end +++++

 There are various ways to define numerical models for fluid forces in under-resolved regions. We show here one possible example. 
Under an assumption of homogeneous solid structure in the under-resolved porous region, the viscous force in the under-resolved PM region is computed using $K_0$ and $K_r^{\alpha}$ as,
\begin{eqnarray}
\label{resistance_form}
\vec{F}^{\alpha}_{PM_{vis}} = - \frac{\nu_{\alpha}}{K_0 K^{\alpha}_r} \rho_{\alpha} \vec{u}_F.
\end{eqnarray}
Also, $K_0$ and $K_r^{\alpha}$  are functions of porosity $\phi$ and density ratio of fluid components, $S_w = \rho_w / \left( \rho_w + \rho_o \right)$.

%% From Here

On the other hand,  an example for a definition of capillary force $\vec{F}^{\sigma}_{PM_{cap}}$ can be written as,
\begin{eqnarray}
\label{F_cap_form}
\vec{F}^{\alpha}_{PM_{cap}} = - \frac{2 \sigma \cos \theta \cdot \mathcal{J}}{\sqrt{K_0 K_r^{\alpha}/ \phi}}  
\frac{ \hat{\nabla} \rho_{o} - \hat{\nabla} \rho_{w}  }{2}  \cdot H \left( At, \left| \partial_x \left( At \right) \right| \right),
\end{eqnarray}
where the hat notation indicates the unit vector and  $\theta$ is the contact angle of the under-resolved porous solid. 
Here $\mathcal{J}$ is the Leverett J-function, defined as $P_c \sqrt{K_0 K^{\alpha}_r/ \phi}/ \sigma \cos \theta$, that is the normalized capillary function of  $S_w$ and $\phi$. 
The Atwood number, $At$, is defined as $At=\left( \rho_{w} - \rho_{o} \right)/ \left( \rho_{w}+\rho_{o} \right)$.
A functional $H$ is a switch function depending on the local multi-component interface condition. 
 This switch function is necessary for the diffusive multi-component model because its non-zero interface thickness may cause excessive artificial force. Moreover, this definition cannot cover a scenario where a component fluid is confined in an under-resolved cell. In order to mitigate this problem, an additional model can be implemented. 
For example, the drainage and stagnation of residual components in a computational cell under certain local pressure and $S_w$  is controlled by referring the Leverett J-function.

The wettability originated from solid parts in the PM site can be taken into account by simply extending  Eq.~(\ref{eq:wall_interaction}). Specifically, $\rho^{'}_{\beta}$ is constructed by  a wall potential from the porous solid, $\rho^{sPM}_{\beta}$, and the fluid density $\rho_{\beta}$ with the ratio of $\phi$ as,
\begin{eqnarray}
\rho^{'}_{\beta} = \phi  \rho_{\beta} + \left( 1- \phi \right) \rho^{sPM}_{\beta},
\end{eqnarray}
where no adjacent regular solid, the solid in the large-scale, exists.
This natural extension for the PM model is one of motivation  to employ the LBM in this study. 
Also, in a computational cell having small porosity, the wettability and friction effects from the regular solid may be suppressed.
It is because fluids are too confined to be influenced by the adjacent regular solids and also wettability and friction effects are already taken into account in the PM region via the input $K_0$, $K_r$, and $P_C$. 
Accordingly, in a cell of a small porosity, wettability and friction effects from adjacent regular solids are switched off.

%%%%%%%%%%%%%%%%%%%%%%%%%%%%%%%%%%%%%%%%%%
%%%%%%%%%%%%%%%%%%%%%%%%%%%%%%%%%%%%%%%%%%
%%        Validation
%%%%%%%%%%%%%%%%%%%%%%%%%%%%%%%%%%%%%%%%%%
%%%%%%%%%%%%%%%%%%%%%%%%%%%%%%%%%%%%%%%%%%
%
\section{Validation}
\label{sec:validation}

\begin{figure*}[htbp]
  \begin{center}
    \includegraphics[clip, width=6 cm]{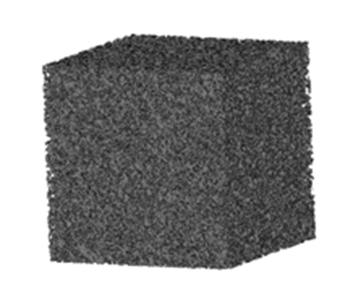}
    \caption{A sampled porous structure applied for the multi-scale simulation in Section~\ref{sec:validation}.}
    \label{fig:typrock_crosssec}
  \end{center}
\end{figure*}

The numerical models and workflow introduced in Sec.~\ref{sec:LB_formalism} and Sec.~\ref{sec:LB_formalism_PM} are validated through a set of benchmark test cases for the single-component as well as multi-component fluid flows.
 As a sampled geometrical model in this section, the PM model shown in Fig.~\ref{fig:typrock_crosssec} has been modified from an open source data \cite{2016_Bultreys}.
The domain size is $256 \times 256 \times 256$ and resolution is 1 $\mu m / pixel$.
The global porosity, $\phi_{glb}$, is $38 \%$.
Firstly, we simulate several cases in this system in order to have inputs for  the following multi-scale simulations. 
% Single-component flow setting
In  the simulation for computing $K_0 \left( \phi \right)$, the domain is mirrored and periodic boundaries are assigned in the flow direction. 
Then gravity $g$ is assigned as driving force. 
The other domain edges are bounded by soild walls. 
The value of viscosity $\nu$ and $g$ is set as $\nu=0.012$ and $1.0 \times 10^{-4}$. 
We evaluate $K_{0}$ as $\phi_{glb} <u> \nu /g$ where $<u>$ is the spatial averaged fluid velocity. 
This $K_{0}$ computation is individually performed for eight-cubes domain which is equally divided from the original domain.
% Pc setting
In  the simulation for computing $P_C \left( S_w \right)$, the oil initially fills the entire domain of the original geometry plus oil reservoir of the top 14-lattices layer. 
The main component on the top/bottom pressure boundary is set as oil/water. 
Their pressure difference $\Delta P$  is initially set as a high value and timely controlled while fixing the pressure in the bottom as $7.33 \times 10^{-2}$.  
Specifically, if the temporal variation of $S_w$ becomes below a certain value, a controller judges as the steady state and shift $\Delta P$ to a next level.
The PM and side walls surrounding the PM are assumed to be water-wet  with a contact angle of 10 degree.
Viscosities for both fluid components are set  to $\nu_w = \nu_o =1.66 \times 10^{-3}$.
% Kr setting
In  the simulation for computing the relative permeability for water and oil, $K_{rw} \left( S_w \right) $ and $K_{ro} \left( S_w \right) $, the domain is mirrored and periodic boundaries are assigned in the flow direction. 
Gravity $g$ is assigned as the driving force. 
Initially, besides the oil mainly occupying the domain,  a small amount of water is distributed in the small-scale PM.
A controller  program adjusts the gravity so that the target capillary number, $Ca = 1.0 \times 10^{-6}$, is achieved at the steady state.
Once the flow reaches the targeted steady state, using  a mass-sink-source (MSS) function \cite{2014_MSS}, the water is injected locally with criteria of local $At$ and velocity fields.
Once $S_w$ reaches the targeted level, the MSS is switched off and the controller adjusts the gravity for  the targeted $Ca$ again. 
This loop is iterated until the relative permeability for all $S_w$ is evaluated.
Wettability conditions are set as  in the $P_C$ simulation.
Viscosities for both components are set as $\nu_w = \nu_o =3.33 \times 10^{-3}$.

%+++++ Figure (Result: Cross section typical PM) ++++++
%+++++ Figure end +++++

\begin{figure*}[htbp]
  \begin{center}
    \includegraphics[clip, width=15 cm]{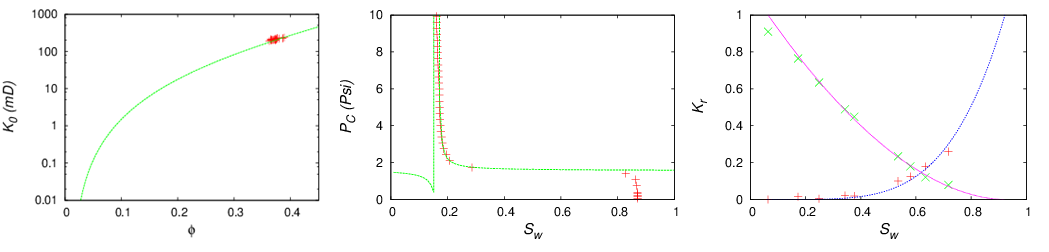}
    \caption{Simulated absolute permeability vs porosity (left), the capillary pressure vs water saturation $S_w$(center), and the relative permeability vs $S_w$ (right) in the PM of Fig.~\ref{fig:typrock_crosssec}. Their fitted results are plotted with lines.}
    \label{fig:Fit_result}
  \end{center}
\end{figure*}

The simulated $K_0 \left( \phi \right)$, $P_C \left( S_w \right)$, $K_{rw} \left( S_w \right) $, and $K_{ro} \left( S_w \right) $  relationships, are fitted with the Kozney-Carman equation, Thommer model, and the Corey model, respectively;
\begin{eqnarray}
\label{Kozney-Carman}
K_0 \left( \phi \right) =\frac{D \left( \phi - \phi_p\right)^2}{72 \tau_p^2 \left(1 - \left( \phi - \phi_p \right) \right)^2} ,   \\
\label{Thommer_model}
P_c \left( S_w \right) =P^{*}_c  \cdot \exp \left(  \frac{\Delta P^{*}_C}{\ln \left(S_w / S_{wref} \right) }\right), \\
\label{Corey model1}
K_{rw}  \left( S_w \right) = \left( \frac{S_w - S_{wi}}{1 -S_{wi}-S_{or} } \right)^{n_w}, \\
\label{Corey model2}
K_{ro}  \left( S_w \right) = \left( \frac{1-S_w - S_{or}}{1 -S_{wi}-S_{or} } \right)^{n_o}.
\end{eqnarray}
 Results for the simulation points and fitting curves are shown next and Fig.~\ref{fig:Fit_result};
\begin{align}
D &=& 801800 \: [ \mbox{mD}],  & &
\tau_p &=&2.5,   & &
\phi_p &=& 0.01, \nonumber \\
 P^{*}_c  &=& 1.515 \: [\mbox{Psi}],  & & 
 \Delta P^{*}_C  &=& 0.0831,  & &
  S_{wref} &=& 0.161, \nonumber \\
   S_{wi} &=& 0.065,  & &
   S_{or} &=& 0.07793,  & &
   n_w &=& 4.408, \nonumber  \\
   n_o&=& 1.844.  & &   & &
\end{align}

Henceforth, unless specifically mentioned, they are used as a standard input for the PM regions modeling in the multi-scale simulation.

%+++++ Figure (Result: Fitting_standard input) ++++++
%+++++ Figure end +++++

%%%
\subsection{Force balance check in single-component fluid flow through porous media}
\label{1D}

The modeled resistance force from PM in Eq.~(\ref{resistance_form}) is validated by checking force balance in single-component fluid flow through a PM region. 
In an arbitrary small domain bounded by periodic boundaries, the PM model is applied everywhere while the gravity $g$ is assigned. The expected force balance can be formulated as $\rho r \cdot \left( \phi u \right) = \rho g$ where $r$ is the resistivity from the PM model, $\phi$ is the porosity, and $u$ is the fluid velocity.
Table~\ref{tab:1d} shows resulted $r \cdot  \left( \phi u \right)$ at various options of viscosity $\nu$, $r$, and $g$.
They agree with input $g$ very well. 
Remembering the formulation of $ r= \nu /K$ derived from Eq.~(\ref{resistance_form}) and definition of $K_0=\phi u \nu /g$, we see that this force balance also indicates the consistence between input $K_0$, $K_{0, in}$, and output $K_0$, $K_{0, out}$.
This is because $K_{0, in}=\nu / r =\phi u \nu   /g= K_{0, out}$ where the formulation of $r$ is used in the first equation and  the force balance is used in the second equation.

\begin{table}[htbp]
    \caption{Results of force balance check in the gravity driving flow through PM}
    \label{tab:1d}
    \begin{center}
    \begin{tabular}{cccc}
        \hline
        $\nu$ & r  & $g$   & $r \cdot \left( \phi u \right) $\\ \hline
        $1.67 \times 10^{-3} $  &  $8.32 \times 10^{-3} $ &      $5.00 \times 10^{-4} $           & $5.00 \times 10^{-4} $ \\
        $3.33 \times 10^{-2} $  &  $1.66 \times 10^{-1} $ &      $1.00 \times 10^{-2} $         & $1.00 \times 10^{-2} $ \\
        $1.67 \times 10^{-1} $  &  $8.32 \times 10^{-1} $ &      $5.00 \times 10^{-2} $          & $5.00 \times 10^{-2} $\\ \hline
    \end{tabular}
    \end{center}
\end{table}

%%%%%%
\subsection{Single-component fluid flow through muti-type porous media}

\begin{figure}[hbtp]
    \centering
    \includegraphics[width=0.7\textwidth]{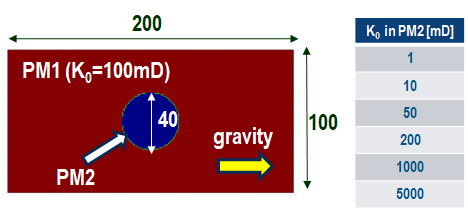}
        \caption{Setups of the the single component fluid flow simulation through various typed porous media.}
        \label{fig:setup_k0cylinder}
\end{figure}

Two-dimensional single-component fluid flow through spatially varied porous structures is simulated.
In a simulation domain of $200 \times 100$, circular-shaped porous medium PM2, whose diameter is 40, is surrounded by the other typed porous medium PM1 as shown in Fig.~\ref{fig:setup_k0cylinder}. 
Many options of the input permeability for PM2 are tried from 1 mD to 5000 mD while one for PM1 is fixed as 100 mD.
The resolution is assumed to be $31.25 \mu m/pixel$.
Gravity is assigned in  the horizontal direction and its value is low enough to realize the Stokes flow regime.
The viscosity $\nu$ is set as $1.66 \times 10^{-3}$.

\begin{figure*}[htbp]
  \begin{center}
    \includegraphics[clip, width=13 cm]{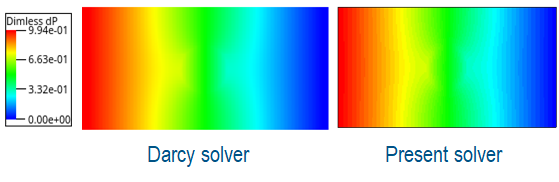}
    \caption{Contours of the pressure profile with the Darcy solver (left) and the present solver (right) in the case of $K_0 = 1 mD$ in PM2. The pressure value is normalized by $F \cdot L_d$ where $F$ is the driving force and $L_d$ is the domain length.}
    \label{fig:k0_cylinder_prs}
  \end{center}
\end{figure*}

The simulated permeability, $K_{0,sim} = \phi_{glb} <u> \nu /g$ where $\phi_{glb}$ is the global porosity, is presented in Table~\ref{tab:model0-pore} together with Darcy solver's results. 
In the Darcy solver, the force balance between driving and resistance term in the Brinkman equation is solved in each definition point \cite{1978_Konikow}. 
Since it ignores the viscous terms and temporal derivative terms, it is applicable only for limited cases in the multi-scale simulation.  
In the present case where the entire domain is covered by porous media, however, the Darcy solver outputs consistent results with the present solver within $0.1 \%$ deviation.
The pressure profiles are compared in Fig.~\ref{fig:k0_cylinder_prs} for the case of $K_0 = 1 mD$ in PM2. The pressure value is normalized by $F \cdot L_d$ where $F$ is the driving force and $L_d$ is the domain length. Its absolute value is shifted to be zero on the right boundary.
Their excellent agreement indicates that the present PM model based on the LBM correctly handles the dynamics in the porous media.

\begin{table}[htbp]
    \caption{Simulated $K_0$ through two-dimensional porous structures using the present and Darcy solver}
    \label{tab:model0-pore}
    \begin{center}
    %\resizebox{\textwidth}{!}{
       \begin{tabular}{| c || ccc |}
           \hline
           $K_0$ in PM2 [mD]                     & $K_{0, sim}$ (Darcy solver)  & $K_{0, sim}$ (Present)  & Deviation  \\ \hline
           1	& 87.52	& 87.45	& 0.08  $\%$\\
           10	& 89.66	& 89.61 & 	0.06  $\%$\\
           50	& 95.83	& 95.8	& 0.03  $\%$\\
           200	& 104.2	& 104.2	& < 0.01 $\%$\\
          1000	& 110.5	& 110.4 & 	0.09  $\%$\\
          5000	& 112.5	& 112.4	&  0.09  $\%$ \\
           \hline
       \end{tabular}
    %}
   \end{center}
\end{table}

%+++++ Figure (Setting: 1D) ++++++
%+++++ Figure end +++++

%%%%%%
\subsection{An imbibition process in one-dimensional porous media}

\begin{figure*}[htbp]
  \begin{center}
    \includegraphics[clip, width=13 cm]{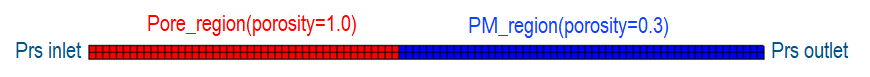}
    \caption{Settings in a one-dimensional PM case. The red and blue color show pore region and PM region of $\phi$=0.3, respectively. In both right and left ends, the pressure boundaries are imposed.}
    \label{fig:1D_setting}
  \end{center}
\end{figure*}

The modeled capillary force from PM in Eq.~(\ref{F_cap_form}) is examined by checking force balance between pressure force and capillary force in one-dimensional PM region and the pore region.
In a simulation domain of 150 lattices, the left-half is set as a pore region and the right-half is set as a PM region of $\phi=0.3$ as shown in Fig.~\ref{fig:1D_setting}, 
In the PM site, besides the standard inputs of $K_0 \left( \phi \right)$, $K_{rw} \left( S_w \right)$, and $K_{ro} \left( S_w \right) $, the input function $P_C \left( S_w \right)$ is set as the constant value of $P_C=0.05$. 
The resolution is assumed to be $4.0 \mu m/pixel$.
The wettability in the PM region is set as water-wet with contact angle 10 degree.
%On both left and right ends of the domain the pressure boundaries are imposed. 
On the right end, the pressure value is set as 0.0733 with $S_w$=0.9995. On the left end, with $S_w$=0.05, the pressure value is set so that pressure difference between both ends, $\Delta P$, is equivalent to 110$\%$ or 90$\%$ of assigned $P_C$ in the PM.  
Initially, oil is mainly filled over the entire domain. Viscosities for both components are $\nu_w = \nu_o =1.66 \times 10^{-3}$.

%+++++ Figure (Setting: 1D) ++++++
%+++++ Figure end +++++

\begin{figure*}[htbp]
  \begin{center}
    \includegraphics[clip, width=13 cm]{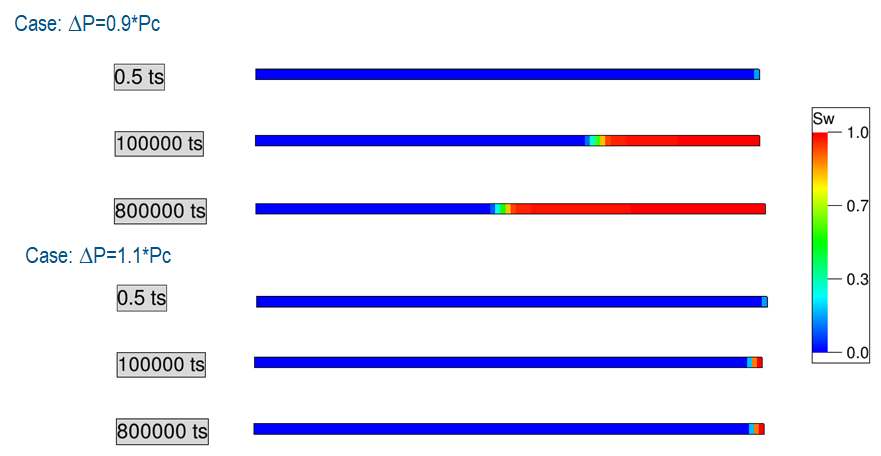}
    \caption{Snapshots of the water distribution at certain timesteps in one-dimensional imbibition processes with the PM model. The top three figures show the case with pressure difference of 90$\%$  assigned $P_C$ in the PM. The bottom three figures show the case with pressure difference of 110$\%$ assigned $P_C$ in the PM.}
    \label{fig:1D_result}
  \end{center}
\end{figure*}

Snapshots of the water distribution at certain timesteps are shown in Fig.~\ref{fig:1D_result}. 
In top and bottom three figures, results in cases with $\Delta P = 0.9 P_C$ and $\Delta P = 1.1 P_C$ are shown, respectively. 
They indicate that the imbibition process can be accurately simulated within 10$\%$ range of the assigned $P_C$ in the PM .

%+++++ Figure (Result: 1D case) ++++++
%+++++ Figure end +++++

%%%%%%
\subsection{An imbibition process in two-dimensional layered channels}

\begin{figure*}[htbp]
  \begin{center}
    \includegraphics[clip, width=13 cm]{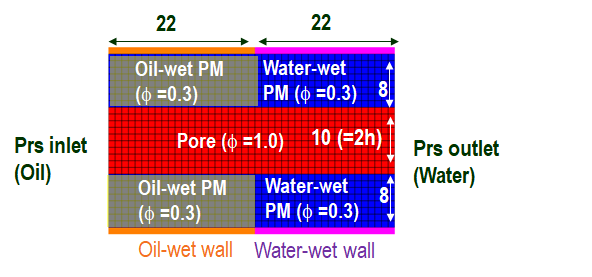}
    \caption{Settings in two-dimensional layered channels. The red, blue, and grey color show pore region, the water-wet and oil-wet porous medium of $\phi$=0.3, respectively.}
    \label{fig:2D_layer_channel_setting}
  \end{center}
\end{figure*}

A typical sequential imbibition process into pores and PM is examined in two dimensional layered channels shown in Fig.~\ref{fig:2D_layer_channel_setting}.
In the domain of $44 \times 26$, the center region, colored red, is pore and regarded as the main channel. 
The resolution is assumed to be $31.25 \mu m/pixel$.
In both sides of the main channel, there are the two different-typed PM of $\phi=0.3$, colored blue and grey.
The left/right PM is oil/water-wet of contact angle 170/10 degree, respectively. 
On top or bottom edge, they are bounded by oil and water-wet walls of the same contact angles as PM. 

The imbibition process is started with sufficiently high pressure difference between inlet and outlet, $\Delta P$. As time goes, $\Delta P$ is gradually reduced.
Due to the scale difference between pores and PM, imbibition into the water-wet PM typically occurs at first once $\Delta P$ becomes sufficiently low. 
When $\Delta P$ is decreased further and becomes comparable with the capillary pressure in the water-wet pores, water invades such pores. 
Later on, as $\Delta P$ is decreased, in contrast to the water-wet scenario, water invades the oil-wet pore at first and the oil-wet PM lastly.
One of main motivations in this section is to capture this sequential process quantitatively.

Besides the standard inputs of $K_0 \left( \phi \right)$, $K_{rw} \left( S_w \right)$, and $K_{ro} \left( S_w \right) $ for the PM region, the input function of $P_C \left( S_w \right)$ is set as the constant value of $P_{C (wwet, PM)}=0.02$ for the water-wet PM and  $P_{C (owet, PM)}=-0.02$ for the oil-wet PM. 
%The domain is bounded by pressure boundaries on the left and right ends.   
%
The pressure value on the right boundary is set as 0.0733 with $S_w$=0.9995. 
According to the Laplace law, the capillary pressure in the main channel is expected to be $P_{C (wwet, pore)}= \sigma \cos \left( 10 ^{\circ} \right)/h=4.92 \times 10^{-3}$ for the water-wet pore and  $P_{C (owet, pore)}= \sigma \cos \left( 170 ^{\circ} \right)/h=-4.92 \times 10^{-3}$ for the oil-wet pore where $h=5$ is the half channel height and $\sigma=0.025$ is the surface tension. 
Considering the estimated capillary pressure above, we gradually decrease $\Delta P$ from $0.03$ to $-0.03$ by changing the pressure value on the left boundary while fixing $S_w$=0.05.
Initially, oil mainly occupies the entire domain. Viscosities for both components are set as $\nu_w = \nu_o =1.66 \times 10^{-3}$.

\begin{figure*}[htbp]
  \begin{center}
    \includegraphics[clip, width=13 cm]{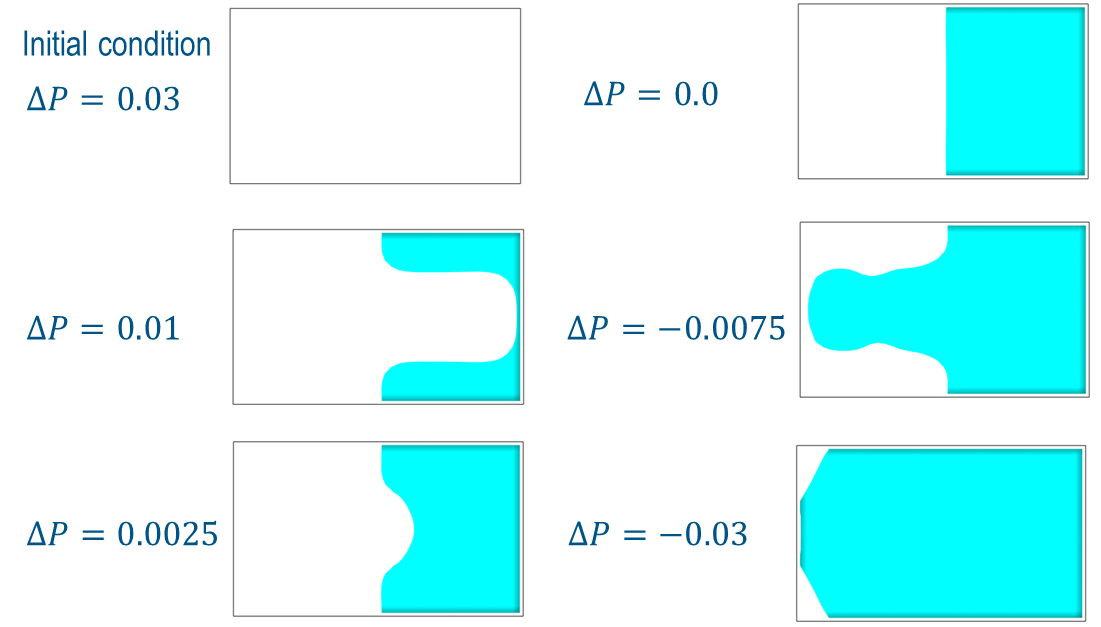}
    \caption{Water iso-surface of $At > 0.5$ at six $\Delta P$ conditions in two-dimensional layered channels.}
    \label{fig:2D_layer_channel_result}
  \end{center}
\end{figure*}

Simulated results are shown in Fig.~\ref{fig:2D_layer_channel_result}. 
The water distributions at six $\Delta P$ conditions are shown with the iso-surface of $At>0.5$.  
At $\Delta P =0.01$ which is below $P_{C (wwet, PM)}$ and above $P_{C (wwet, pore)}$ water invades the water-wet PM.
When $\Delta P =0.0025$ and $\Delta P =0.0$ which are below $P_{C (wwet, pore)}$ and above $P_{C (owet, pore)}$ water invades the water-wet pore.
When $\Delta P =-0.0075$ which are below $P_{C (owet, pore)}$ and above $P_{C (owet, PM)}$ water invades the oil-wet pore.
Lastly, when $\Delta P =-0.03$ which is below  $P_{C (owet, PM)}$, most of entire domain is filled by water.
As a result, the current PM model successfully reproduce the expected sequential imbibition process to pores and PM quantitatively.

%+++++ Figure (Setting 2D_layer_channel_result) ++++++
%+++++ Figure end +++++

%+++++ Figure (Result: 2D_layer_channel_result) ++++++
%+++++ Figure end +++++

%%%%%%
\subsection{Porous media of a large cone-shaped hole}
\label{Model0}

\begin{figure}[!hbtp]
    \centering
        \includegraphics[width=0.7\textwidth]{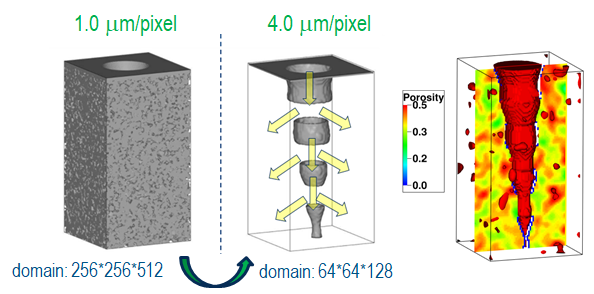}
        \caption{Visualization of grains. Solid surfaces in the resolved-PM case (left) and in the under-resolved-PM case (center). The iso-surface of porosity larger than 0.9 is displayed in the right together with the color contour of the porosity on a center plane.}
        \label{fig:model0-grain}
\end{figure}

Using an in-house designed PM model, that has a large cone-shaped hole bounded by the solid walls and partially connecting to the PM regions, single-component and multi-component fluid flows are simulated using two different resolutions.
Finer resolution, 1 $\mu m /pixel$, allows us to resolve the PM structures fully and capture the geometry shown in the left figure of Fig.~\ref{fig:model0-grain}. 
On the other hand, coarse resolution, 4  $\mu m / pixel$, under-resolves the PM structures but can resolve only large-scale solid walls  on the boundaries between hole and PM as shown in the middle figure of Fig.~\ref{fig:model0-grain}. 
The PM model in Sec.~\ref{sec:LB_formalism_PM} is applied only for the under-resolved regions in the coarse-resolution case, using porosity distribution presented in the right figure of Fig.~\ref{fig:model0-grain}.  % The mean value of porosity in the PM regions is 0.37.
The contributions of the resolved and under-resolved regions to the global porosity is $41 \%$ and $55 \%$, respectively.
In this section, we mainly examine the consistence between such fully-resolved-PM case and under-resolved-PM case.

\begin{figure*}[htbp]
  \begin{center}
    \includegraphics[clip, width=14 cm]{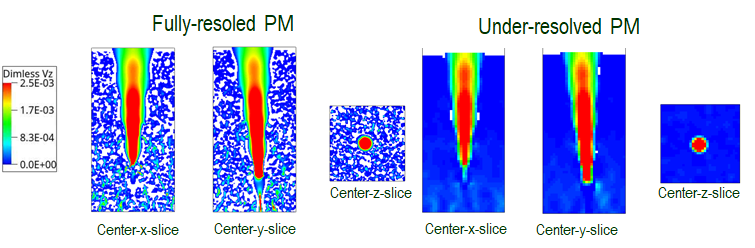}
    \caption{Contour plots of the non-dimensional z-velocity using $g L^2 / \nu$, where $g$ is gravity, $\nu$ is the kinematic viscosity, and $L$ is the characteristic length.}
    \label{fig:model0-v}
  \end{center}
\end{figure*}

% Single-component flow setting
In the simulation for computing $K_0$, the domain and gravity are set in the same manner as the $K_0$ simulation for the PM model in Fig.~\ref{fig:typrock_crosssec}.
The domain sizes in the resolved-PM case and the under-resolved-PM case, before the mirroring, are $256 \times 256 \times 512$ and $64 \times 64 \times 128$, respectively.
The viscosity is set as $\nu=0.166$ for the resolved-PM case and $\nu=0.0166$ for the under-resolved-PM case. 
The gravity, $g$, is set as $1.0 \times 10^{-4}$ for the resolved-PM case and $1.5 \times 10^{-4}$ for the under-resolved-PM case.
% Single-component flow result
Table~\ref{tab:k0-model0} shows computed $K_0$ in the resolved-PM and under-resolved-PM case. 
Although the resolved-PM case requires more than 20-times CPU hours compared to the under-resolved-PM case, their $K_0$ values are consistent within $6.1\%$ deviation. 
The deviation possibly comes from connectivity among the PM cells and improper assignments of input $K_0$ for the PM model on boundaries between the hole and the PM. 
On such boundaries, homogenous PM models shown in Fig.~\ref{fig:typrock_crosssec} may not be accurate. 
Fig. \ref{fig:model0-v} shows comparisons of flow fields on XY-/XZ-/YZ-planes, displayed with the non-dimensionalized z-velocity by $g L^2 / \nu$ where $L$ is the characteristic length. 
It shows that the PM model enables us to capture the reasonable flow field even inside the PM.
According to our original method in which the under-resolved regions are regarded as solid, there is no main flow passages through the domain and therefore $K_0$ results in almost zero. 
The proposed methods and the PM model address this issue effectively and provide accurate $K_0$ and velocity profiles while saving computational costs largely.

\begin{table}[!hbtp]
    \caption{Absolute permeability $K_0$ in porous media of a large cone-shaped hole}
    \label{tab:k0-model0}
    \begin{center}
    \begin{tabular}{ll}
        \hline
        Case                   & $K_0$ (mD) \\ \hline
        Resolved-PM  & 458 \\
        Under-resolved-PM     & 486   \\ \hline
    \end{tabular}
    \end{center}
\end{table}

%+++++ Figure (Result1: Model0) ++++++
%+++++ Figure end +++++

\

\begin{figure*}[htbp]
  \begin{center}
    \includegraphics[clip, width=14 cm]{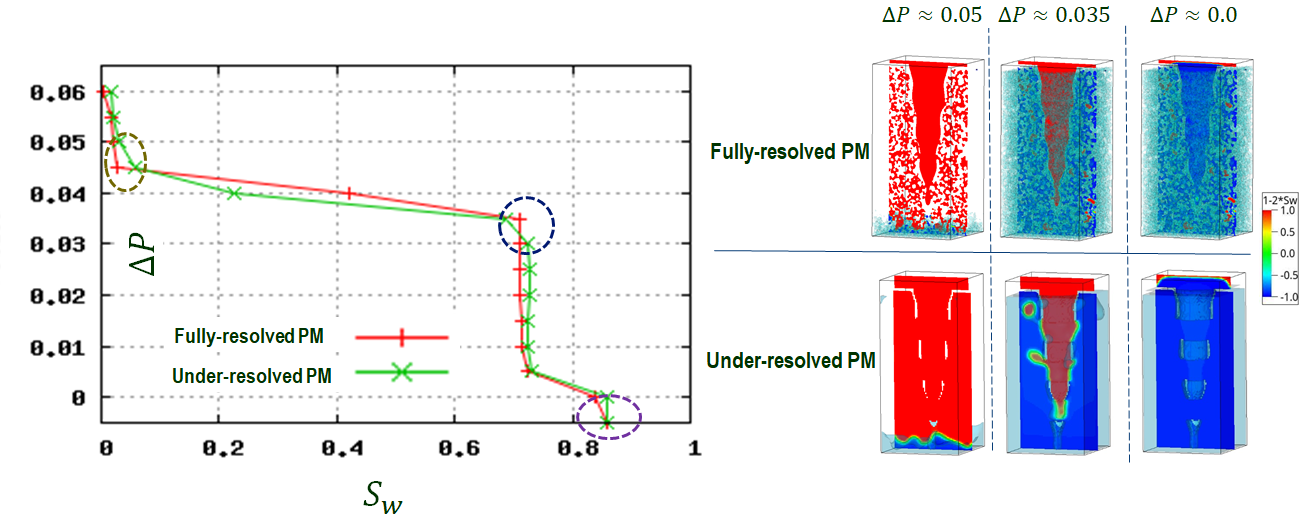}
    \caption{Capillary pressure curve in terms of water-saturation $S_w$ in the resolved-PM case and the under-resolved-PM case  in the left figure. In the right figures, the water distributions at certain stages, marked with dotted circles in the left figure, are shown for both cases using the iso-surface of $At > 0.5$ and color contour of $At$ on the central cutting plane.}
    \label{fig:model0_result}
  \end{center}
\end{figure*}

% Two-component flow setting
In the simulation for computing $P_C$, the domain settings, initial conditions, and simulation processes follow the same manner as the $P_C$ simulation for the PM in Fig.~\ref{fig:typrock_crosssec}.
For the PM settings, the standard input of $K_0 \left( \phi \right)$, $P_C \left( S_w \right)$, $K_{rw} \left( S_w \right) $ and $K_{ro} \left( S_w \right) $ are used with the same wettability condition.
Viscosities for both components are set as $\nu_w = \nu_o =1.66 \times 10^{-3}$.
%  Two-component flow results
The resulted capillary pressure, $\Delta P$, in terms of the water saturation $S_w$ is shown in the left figure of  Fig.~\ref{fig:model0_result}.
Here, a resolution factor 4 is multiplied for $\Delta P$ of the resolved-PM case for fair comparisons in the lattice unit.
The water distributions at certain stages, marked with the dotted circles in the left figures, are shown in the right figures using the light-blue iso-surfaces of $At>0.5$ and color contours on the central cutting plane. 
%+++++ Figure (Result1: Model0) ++++++
%+++++ Figure end +++++
% Discussion
The capillary pressure curves show the main entry pressure around $\Delta P=0.045$ is accurately captured with the PM model compared to the resolved-PM case within 5 $\%$ deviation.
This main entry occurs in the PM region and therefore accuracy of the input $P_C$ curve plays an important role. 
Also, the entry pressure into the large hole around $\Delta P=0.005$ is accurately captured with the PM model.
As seen in the right figures of Fig.~\ref{fig:model0_result}, some oil bubbles are observed in the PM region in the middle of imbibition. 
They are possibly caused by the difficulty to capture the exact steady state.
It is because the dynamics in the PM region is usually very slow due to high viscous force from the complex porous structure and complex invasion paths.
As a result, the simulation controller sometimes insufficiently judges the steady state and proceeds to the next stage. 
Nevertheless, the capillary pressure curves in the under-resolved PM and resolved PM case are reasonably matched while the simulation time is saved by a factor of 30.

%%%%%%%
\subsection{Porous media made from a typical Carbonate rock}

\begin{figure*}[htbp]
  \begin{center}
    \includegraphics[clip, width=10 cm]{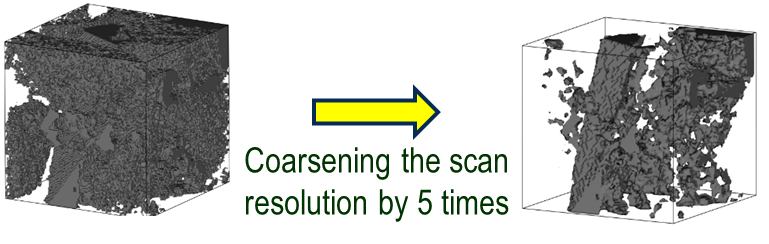}
    \caption{PM structures captured with resolution 0.758 $\mu m / pixel$ (left) and resolution 3.79  $\mu m / pixel$.}
    \label{fig:SAVII_geom}
  \end{center}
\end{figure*}

\begin{figure*}[htbp]
  \begin{center}
    \includegraphics[clip, width=14 cm]{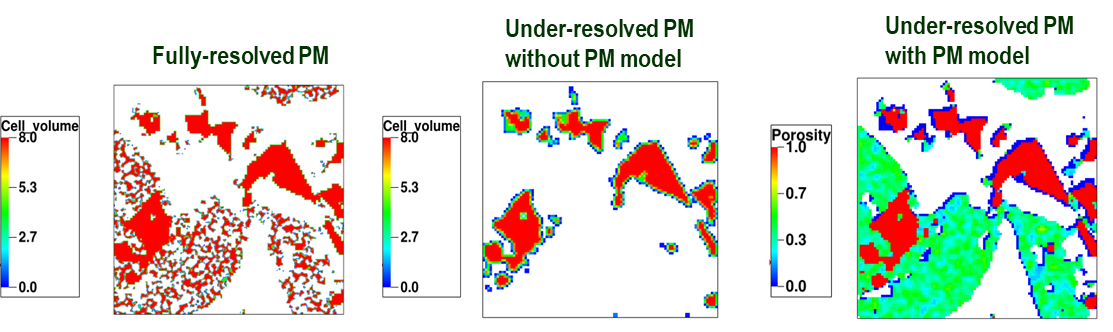}
    \caption{Geometries of simulated fluid cells under three different conditions. The fluid cells in the resolved-PM case with the color contours of the cell volume (left). The fluid cells captured with the coarse resolution where the under-resolved-PM is regarded as the solid (middle). The fluid cells captured with the coarse resolution plus fluid cells handled by the PM model together with the colored porosity distribution. }
    \label{fig:SAVII_crosssection}
  \end{center}
\end{figure*}

Using an in-house designed PM model made from images of a typical Carbonate rock \cite{2016_Bultreys}, single-component and multi-component fluid flows are simulated using two different resolutions.
In order to produce multi-scale structures explicitly, the porous structures in Fig.~\ref{fig:typrock_crosssec} are patched to the original images as small-scale PM structures. 
Specifically, the images of Fig.~\ref{fig:typrock_crosssec} are patched with a scaling so that resolution 0.758 $\mu m / pixel$ allows us to resolve all PM structures. 
As a result, the geometry shown in the left figure of Fig.~\ref{fig:SAVII_geom} is captured at this resolution. 
Then the images are coarsen by 5 times. 
The coarse resolution, 3.79  $\mu m / pixel$, under-resolves small-scale PM but can resolve only large-scale PM structures shown in the right figure of Fig.~\ref{fig:SAVII_geom}. 
The contributions of the resolved and under-resolved regions to the global porosity is $27 \%$ and $19 \%$, respectively.
Originally, the under-resolved PM regions are treated as solid in the simulation, but now it can be handled by the PM model.
In the sense of clarification, geometries of fluid cells on a certain cross section are presented for three compared conditions in Fig.~\ref{fig:SAVII_crosssection}.
The left figure shows the fluid cells in the resolved-PM case with color contours of fluid volume.
The middle figure shows the fluid cells captured with the coarse resolution where the under-resolved PM is regarded as solid. 
It indicates that many small-scale structures are missed compared to the resolved-PM case.
The right figure shows that the fluid cells captured with the coarse resolution plus the fluid cells handled by the PM model using the colored porosity distribution. It explicitly shows the connectivity among large-scale PM are enhanced compared to the middle figure.

%+++++ Figure (SAVII geom) ++++++
%+++++ Figure end +++++

%+++++ Figure (SAVII geom) ++++++
%+++++ Figure end +++++

\begin{figure*}[htbp]
  \begin{center}
    \includegraphics[clip, width=14 cm]{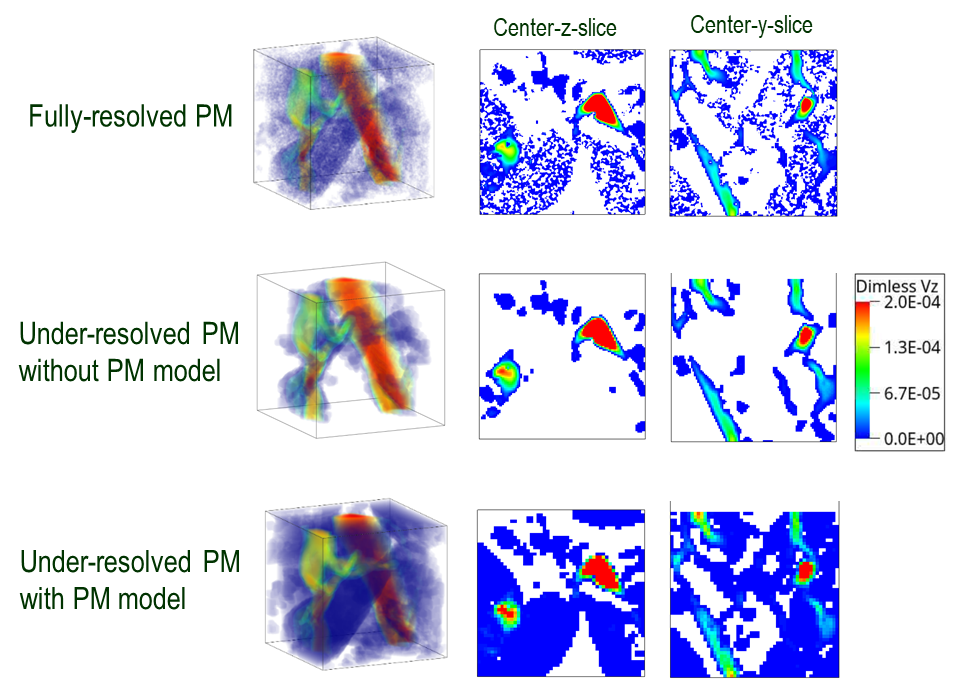}
    \caption{Dimensionless velocity profiles over the domain (left column) and on the center z- (center column) and y- (right column) slices. Velocity is non-dimensionalized by $g L^2 / \nu$. The resolved PM case (top raw), the under-resolved-PM case without the PM model (middle raw), and the under-resolved-PM case with the PM model (bottom raw) are compared.}
    \label{SAVII_K0vel}
  \end{center}
\end{figure*}

% Single-component flow setting
In the simulation for computing $K_0$, the domain and gravity are set in the same manner as the simulation for the PM model in Fig.~\ref{fig:typrock_crosssec}.
The domain sizes in the resolved-PM case and the under-resolved-PM case are $500 \times 500 \times 500$ and $100 \times 100 \times 100$, respectively.
The viscosity is set as $\nu=0.166$ for the resolved-PM case, $\nu=0.0166$ for the under-resolved-PM case without the PM model, and $\nu=0.012$ for the under-resolved-PM case with the PM model. The gravity, $g$, is set as $5.7 \times 10^{-6}$ for the resolved-PM case and $1.4 \times 10^{-3}$ and $1.4 \times 10^{-5}$ for two under-resolved-PM case.
% Single-component flow result
Table~\ref{tab:k0-SAVII} shows $K_0$ computed in the resolved-PM and under-resolved-PM cases. 
In contrast to the case in Section~\ref{Model0}, the under-resolved-PM case without the PM model outputs comparable $K_0$ to one in the resolved-PM case. 
This is because main flow passages through the domain exist and largely contribute to $K_0$.  
Qualitatively, due to less connectivity of large-scale PM, $K_0$ is slightly reduced from the resolved-PM case. 
On the other hand, the PM model enhanced their connectivity and results in slightly higher $K_0$. 
All of three cases shows almost comparable $K_0$ within $6 \%$ deviation and consistent velocity profiles in Fig.~\ref{SAVII_K0vel}.
This fact demonstrates the PM model works properly.
Moreover, the computational cost for simulation is saved from resolved-PM to the under-resolved-PM case with the PM model by factor of 5.

\begin{table}[!hbtp]
    \caption{Absolute permeability $K_0$ in PM made from a typical Carbonate rock}
    \label{tab:k0-SAVII}
    \begin{center}
    \begin{tabular}{ll}
        \hline
        Case                   & $K_0$ (mD) \\ \hline
        Resolved-PM  & 2121 \\
        Under-resolved-PM without the PM model    & 2040   \\ 
        Under-resolved-PM with the PM model    & 2279   \\  \hline
    \end{tabular}
    \end{center}
\end{table}

%+++++ Figure (SAVII k0 result) ++++++
%+++++ Figure end +++++

\begin{figure*}[htbp]
  \begin{center}
    \includegraphics[clip, width=13 cm]{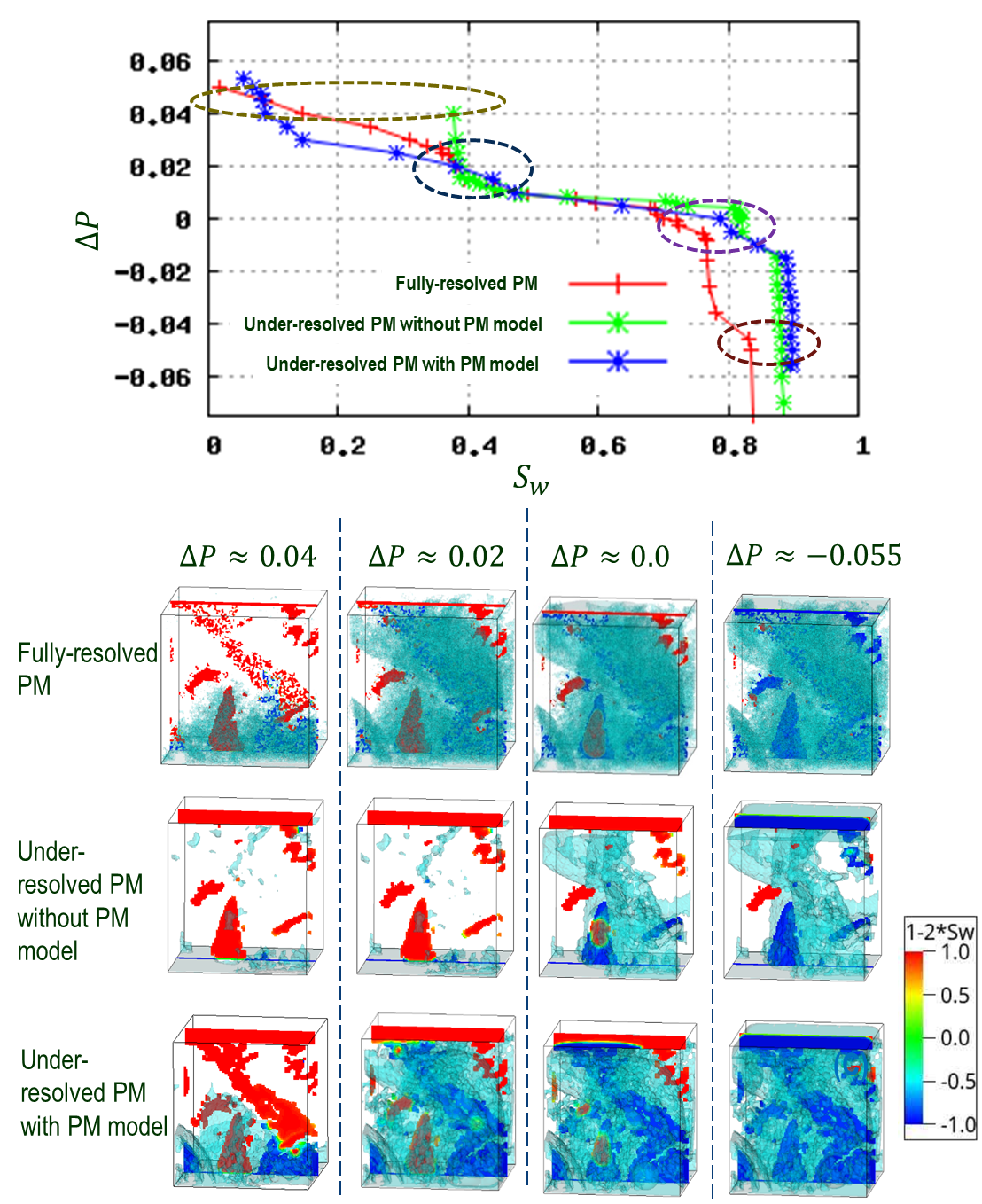}
    \caption{Capillary pressure curve in terms of water-saturation $S_w$ in the resolved-PM case and the under-resolved-PM case  with/without the PM model in the top figure. In the bottom figures, the water distributions at certain stages, marked with the dotted circles in the top figure, are shown for all of cases using the iso-surface of $At > 0.5$ and color contour of $At$ on the central cutting plane.}
    \label{fig:SAVII_pcresult}
  \end{center}
\end{figure*}

\begin{figure*}[htbp]
  \begin{center}
    \includegraphics[clip, width=10 cm]{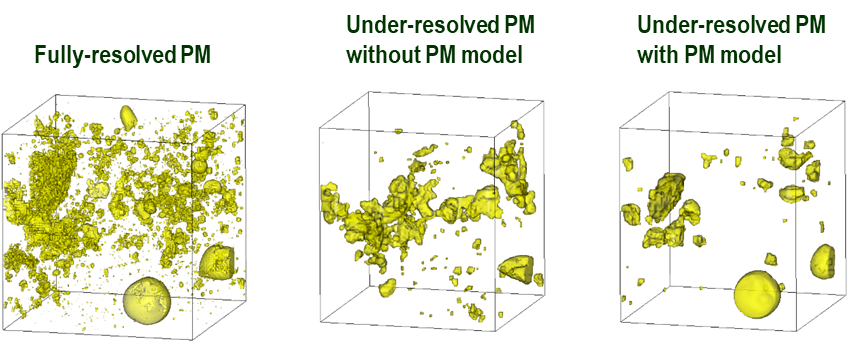}
    \caption{Residual oil patterns at the last stage of the imbibition process, $\Delta P \approx -0.055$ displayed by the iso-surface of $At<-0.5$.}
    \label{fig:SAVII_residualoil}
  \end{center}
\end{figure*}

% Two-component flow setting (to be modified)
In the simulation for computing $P_C$, the domain settings, initial conditions, and simulation processes follow the same manner as the simulation for the PM model in Fig.~\ref{fig:typrock_crosssec}.
For the PM settings, the standard input of $K_0 \left( \phi \right)$, $P_C \left( S_w \right)$, $K_{rw} \left( S_w \right) $ and $K_{ro} \left( S_w \right) $ are used with the same wettability condition.
Viscosities for both components are set as $\nu_w = \nu_o =1.66 \times 10^{-3}$.
%  Two-component flow results
The resulted capillary pressure, $\Delta P$, in terms of the water saturation $S_w$ is shown in the top figure of  Fig.~\ref{fig:SAVII_pcresult}.
Here, a resolution factor 5 is multiplied for $\Delta P$ of the resolved-PM case for fair comparisons in the lattice unit.
According to the displayed $S_w$ in the under-resolved-PM case without the PM model, the volume of the under-resolved PM, which is regarded as solid in this case, is assumed to be filled by 89 $\%$ water all time.
The water distributions at certain stages, marked with the dotted circles in the top figure, are shown in the bottom figures using the light-blue iso-surfaces of $At>0.5$ and color contours on the central cutting plane. 
Between $\Delta P \approx 0.04$ and $\Delta P \approx 0.02$, the water mainly invades the small-scale PM regions. 
Without the PM model this stage cannot be simulated. 
The under-resolved PM case with the PM model shows reasonably consistent results with the resolved-PM case.
Between $\Delta P \approx 0.02$ and $\Delta P \approx 0.0$, the water mainly invades the large-scale PM regions. As seen in the Pc curve, all of three cases show excellent agreements at this stage.
%+++++ Figure (SAVII pc result) ++++++
%+++++ Figure end +++++
Between $\Delta P \approx 0.0$ and $\Delta P \approx -0.055$, oil in remainning spaces is washed out. 
At this stage, connectivity among large-scale PM regions play an important role.
At the last stage of the imbibition process, the residual components patterns are shown in Fig.~\ref{fig:SAVII_residualoil} with the iso-surface of $At<-0.5$.
Although it is difficult to compare their detailed structures by looking at the iso-surface at different resolutions, it may be fair to compare large-scale patterns.
Indeed we observe that the under-resolved PM case without the PM model misses some of big oil bulbs in the resolved-PM case but the PM model successfully captures them consistently.
Moreover the computational cost in the under-resolved PM case is saved by a factor of 43 compared to the resolved-PM case.

%+++++ Figure (SAVII residual oil result) ++++++
%+++++ Figure end +++++

%%%%%%%
\subsection{Relative permeability with coarsen images of sampled porous media}

In the simulation computing relative permeability $K_r$ in Fig.~\ref{fig:Fit_result}, the sampled PM introduced in Fig.~\ref{fig:typrock_crosssec} is fully resolved with resolution 0.758 $\mu m / pixel$.
In this section, the images are coarsen by 16 times and then the size of computational domain becomes $16 \times 16 \times 16$  in which the PM structure is under-resolved everywhere.
By performing the $K_r$ simulation in such system using the PM model, we see if the results are comparable with the fully-resolved PM case.
The domain, gravity, and wettability are set in the same manner as the $K_r$ simulation in Fig.~\ref{fig:Fit_result}
The viscosity for both components is set as $3.33 \times 10^{-3}$.

\begin{figure*}[htbp]
  \begin{center}
    \includegraphics[clip, width=15 cm]{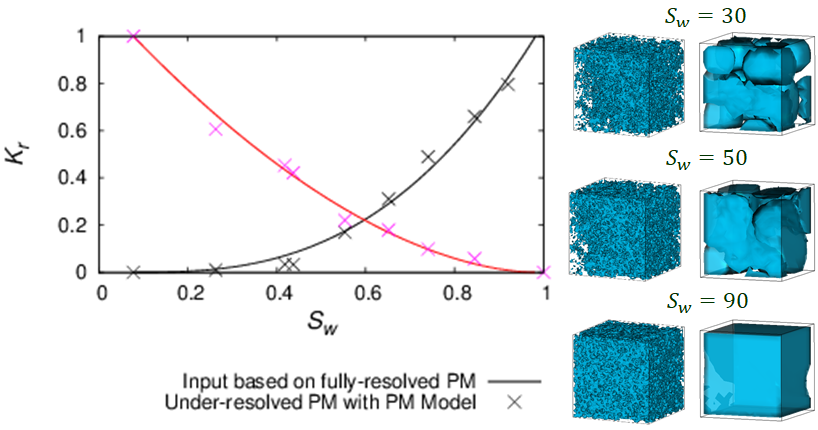}
    \caption{Relative permeability in the under-resolved-PM case with black crosses for $K_{rw}$ and purple crosses for $K_{ro}$. The red and black lines are the input $K_r$ based on the fully-resolved-PM case in Fig.~\ref{fig:Fit_result}. In the right, the water distribution is compared between the fully-resolved PM (left) and under-resolved-PM (right) case at certain $S_w$.}
    \label{fig:PM0_PM_Kr}
  \end{center}
\end{figure*}

In Fig.~\ref{fig:PM0_PM_Kr}, the resulted $K_r$ curves are compared with the input $K_r$ curves which are from the fully-resolved case in Fig.~\ref{fig:Fit_result}. They show excellent agreements.
Also, in the right images, the water distributions are compared  at certain $S_w$ between the fully-resolved PM case and under-resolved PM case. 
Although it is difficult to compare detailed water distribution under such large resolution difference, we try to show them using the iso-surface of $S_w \cdot \phi > 0.9$ for the fully-resolved PM case and  $S_w \cdot \phi > 0.35$ for the under-resolved PM case. Results show that the water volume grows homogenously in both of cases. 
As a result, the proposed multi-scale approach allows us to perform the consistent $K_r$ simulation in the under-resolved PM with the fully-resolved PM case. 
Due to the coarsen resolution, the simulation time is saved by 200 times from the fully-resolved-PM case. 

%+++++ Figure (PM0 Kr result) ++++++
%+++++ Figure end +++++

%
%%%%%%%%%%%%%%%%%%%%
\section{Summary}
\label{sec:summary}
%%%%%%%%%%%%%%%%%%%%

Computational models and a workflow for efficient multi-component-flow simulations in multi-scale solid structures are proposed and validated through a set of benchmark test cases.
Specifically, using pre-computed physical properties such as the absolute permeability $K_0$, capillary pressure $P_C$, and relative permeability $K_r$ from resolved simulations in tiny subdomains of the representative porous structures with fine resolution, local fluid force is constructed to account for viscous, capillary, and pressure forces from under-resolved porous media (PM), as well as other local PM and fluid information such as the porosity and water saturation.
In this way, flow simulation in multi-scale solid structures becomes feasible with practical resolution.
The validation is conducted by comparing with analytic solutions and computed results with much finer-resolution corresponding cases resolving the PM structures.
In addition to artificially established systems, the tested benchmarks include in-house designed models for multi-scale complex porous structures.

% Result
In the $K_0$ simulation, comparing with analytic solutions, it is confirmed that the PM model works accurately for viscous force as expected. 
It shows the consistence with the Darcy solver within $1 \%$ accuracy in the multi-type PM case.
Also, compared with the resolved-PM case,  $K_0$ in the under-resolved-PM case is consistent within $6 \%$ accuracy in the in-house designed PM case.
The simulation cost is saved by a factor of 20 at best.
In the $P_C$ simulation, comparing with analytic solution, it is confirmed that the PM model can accurately reproduce the capillary force, and can also quantitatively reproduce the typical sequential imbibition process in PM and pores.
Moreover, in the in-house designed PM case, the entry pressures to both PM and pores agree well between the under-resolved-PM case and the resolved-PM case.
The PM model successfully captures the major residual oil blobs at the last stage of the imbibition process consistently with the resolved-PM case.
The simulation cost is saved by a factor of 43 at best.
In the $K_r$ simulation, comparing the fully-resolved-PM case, the PM model can accurately reporoduce the relative permeability and water distributions in the under-resolved PM.
The simulation cost is saved by a factor of 200 at best.

Further applications of this under-resolved-simulation approach for a reservoir rock can be found in \cite{2021_AFager}.
The models and methodology in this study can be extensively applied for various engineering systems of multi-scale porous structures such as exampled cases in the introduction. Accordingly, further explorations and developments are expected in near future.

\end{document}